\documentclass[aps,prl,twocolumn,superscriptaddress,nofootinbib,floatfix]{revtex4}

\usepackage{amsmath,amssymb,bm,graphicx}
\usepackage[colorlinks=true,citecolor=blue,urlcolor=blue]{hyperref}
\usepackage{microtype}
\usepackage{comment}

\newcommand{\reff}{r_{\rm eff}}

\newcommand{\dd}{\mathrm{d}}

\newcommand{\vs}{\vspace*}

\begin{document}
\bibliographystyle{apsrev}

\title{
Hidden Perturbative Structure in Attosecond Streaking}

\author{Jingsong Gao}
\affiliation{J. R. Macdonald Laboratory, Department of Physics, Kansas State University, Manhattan, Kansas 66506, USA}
\author{Mahmudul Hasan}
\affiliation{J. R. Macdonald Laboratory, Department of Physics, Kansas State University, Manhattan, Kansas 66506, USA}
\author{Zach Eisenhutt}
\affiliation{J. R. Macdonald Laboratory, Department of Physics, Kansas State University, Manhattan, Kansas 66506, USA}
\author{Vladislav~V.~Serov}
\affiliation{Department of Medical Physics, Saratov State University,
             Saratov 410012, Russia}
\author{Hao Liang}
\affiliation{Max-Planck-Institut f\"ur Kernphysik, Heidelberg 69117, Germany}
\author{Meng Han}
\email{meng9@ksu.edu}
\affiliation{J. R. Macdonald Laboratory, Department of Physics, Kansas State University, Manhattan, Kansas 66506, USA}
\author{Anatoli S. Kheifets}
\email{A.Kheifets@anu.edu.au}
\affiliation{Research School of Physics, The Australian National University, Canberra ACT 2601, Australia}

\date{\today}

\begin{abstract}
Attosecond streaking is inherently multiphoton, yet we uncover a
hidden perturbative structure revealing the underlying
continuum--continuum angular-momentum propensity. By filtering the
inversion-odd response in helium streaking data at the IR fundamental,
we isolate an angular distribution dominated by the $P_1$ and $P_3$
Legendre components, whose complex coefficients yield an effective
$s/d$ two-photon amplitude ratio. Multichannel, two-electron
time-dependent Schr\"odinger equation calculations demonstrate that
the characteristic Fano propensity survives electron correlation.
Incorporating the experimental XUV attochirp reproduces the measured
emission-edge trend, while near-threshold calculations predict
resonant amplitude and phase signatures associated with the bound He
$np$ Rydberg series. This symmetry- and frequency-filtering approach
offers a novel spectroscopic framework with potential applications
to atomic, molecular, and solid-state systems where attosecond
streaking is established.
\end{abstract}

\maketitle

Attosecond streaking is commonly conceptualised as a direct mapping of
photoelectron emission time onto final momentum by a phase-locked
infrared (IR)
field~\cite{Hentschel2001,Itatani2002,Kitzler2002,Kienberger2004,Sansone2006,Schultze2010}. 
However, the familiar semiclassical relation between the photoelectron
momentum shift and the streaking-field vector potential evaluated at
delay $\tau$, $\Delta p \simeq -A(\tau)$, conceals a richer quantum
process: the measured signal arises from interference among
photoelectron pathways coupled by the IR
field~\cite{Dahlstrom2013}. Not only does this coupling redistribute 
energy but also orbital angular momentum, rendering angle-resolved
streaking intrinsically multiphoton. Recent helium measurements
exposed this complexity directly: at a modest IR intensity of $4.4 \times
10^{11}~\mathrm{W/cm}^2$, multiple IR-photon orders populate
angular-momentum channels far beyond those accessible through a single
IR-photon transition~\cite{Gao2026NP}. Their numerical solution
of the time-dependent Schr\"odinger equation (TDSE) driven by
reconstructed experimental fields reproduce the measured energy- and
angle-resolved traces with exceptional fidelity.

A fundamental guide to this angular redistribution is Fano's
propensity rule, which dictates that photon absorption preferentially
increases orbital angular momentum while stimulated emission decreases
it~\cite{Fano1985,Busto2019}. For a $p$-wave continuum generated in
helium by an XUV pulse and dressed by a single IR photon, this
continuum--continuum (CC) propensity strongly favors $p \to d$
absorption and $p \to s$ stimulated emission. This angular-momentum
competition is central to angle-resolved reconstruction of attosecond
beating by interference of two-photon transitions
(RABBITT)~\cite{Heuser2016,Ivanov2017,Bray2018b,Cirelli2018,Busto2019,Joseph2020}
as well as dichroic retrievals of two-photon amplitudes in circular
RABBITT~\cite{Han2023NatPhys,Han2023Optica,Kheifets2024,Serov2026},
both of which enable protocols to extract partial-wave ratios and
quantitatively test Fano's propensity rule.

The weak IR fields typically employed in RABBITT naturally
restrict this process to a perturbative regime, greatly simplifying
the isolation and analysis of individual angular-momentum channels.
Whether a comparable analysis remains viable in attosecond streaking,
where a stronger laser field drives intrinsically multiphoton
dynamics, is far from obvious. Here we build on the heterodyne picture
of streaking~\cite{Dahlstrom2013}, in which the fundamental-frequency
modulation arises from interference between the direct XUV amplitude
and the one-IR-photon-assisted amplitudes. We demonstrate that the
inversion-odd response yields a clean perturbative angular-momentum
structure that can be disentangled from the nonperturbative streaking
signal and mapped onto the $s/d$ continuum--continuum propensity. The
isolated signal exhibits an angular distribution dominated by the
$P_1$ and $P_3$ Legendre polynomials. Crucially, the upper and lower
energy edges of the spectrogram are highly selective, preferentially
isolating single-IR-photon absorption and stimulated emission
pathways, respectively. Furthermore, unlike RABBITT, this formulation
operates directly on an isolated-pulse streaking spectrogram, entirely
bypassing the need for a phase-coherent train of XUV harmonics.

Although this symmetry- and frequency-selection principle generalizes
beyond helium, the specific reduction to a $P_1+P_3$ angular structure
and its direct mapping onto an $s/d$ amplitude ratio rest on the
assumption of an initial $s$ state and parallel linear laser
polarizations. Other initial configurations involve a different set of
angular-momentum channels; thus, for simplicity, we focus our
presentation below on helium. In our proof-of-principle time-dependent
Schr\"odinger equation (TDSE) simulations, we drive helium with a
transform-limited Gaussian XUV pulse, tuning its central frequency
across a broad spectral range. The effective $s/d$ amplitude ratio,
extracted from the relative magnitude and phase of the $P_1$ and $P_3$
coefficients, closely tracks the analytical prediction for a
perturbative Coulomb continuum--continuum transition. Moving beyond
this idealized benchmark, we incorporate the experimentally
reconstructed XUV attochirp into our TDSE framework. The resulting
effective $s/d$ ratio exhibits excellent agreement with experimental
streaking data measured on the emission edge of the photoelectron
momentum distribution. Furthermore, upon approaching the ionization
threshold, the TDSE uncovers a pronounced resonant structure
governed by the discrete He $np$ Rydberg series. Consequently, even
within the highly nonperturbative regime of attosecond streaking, a
carefully selected observable preserves an elegant connection to
underlying single-photon dynamics, establishing a novel spectroscopic
pathway to probe continuum propensity and bound-state resonances.

\begin{figure}[t]
\centering
\includegraphics[
  width=\columnwidth,
  trim=0 52bp 0 107bp,
  clip
]{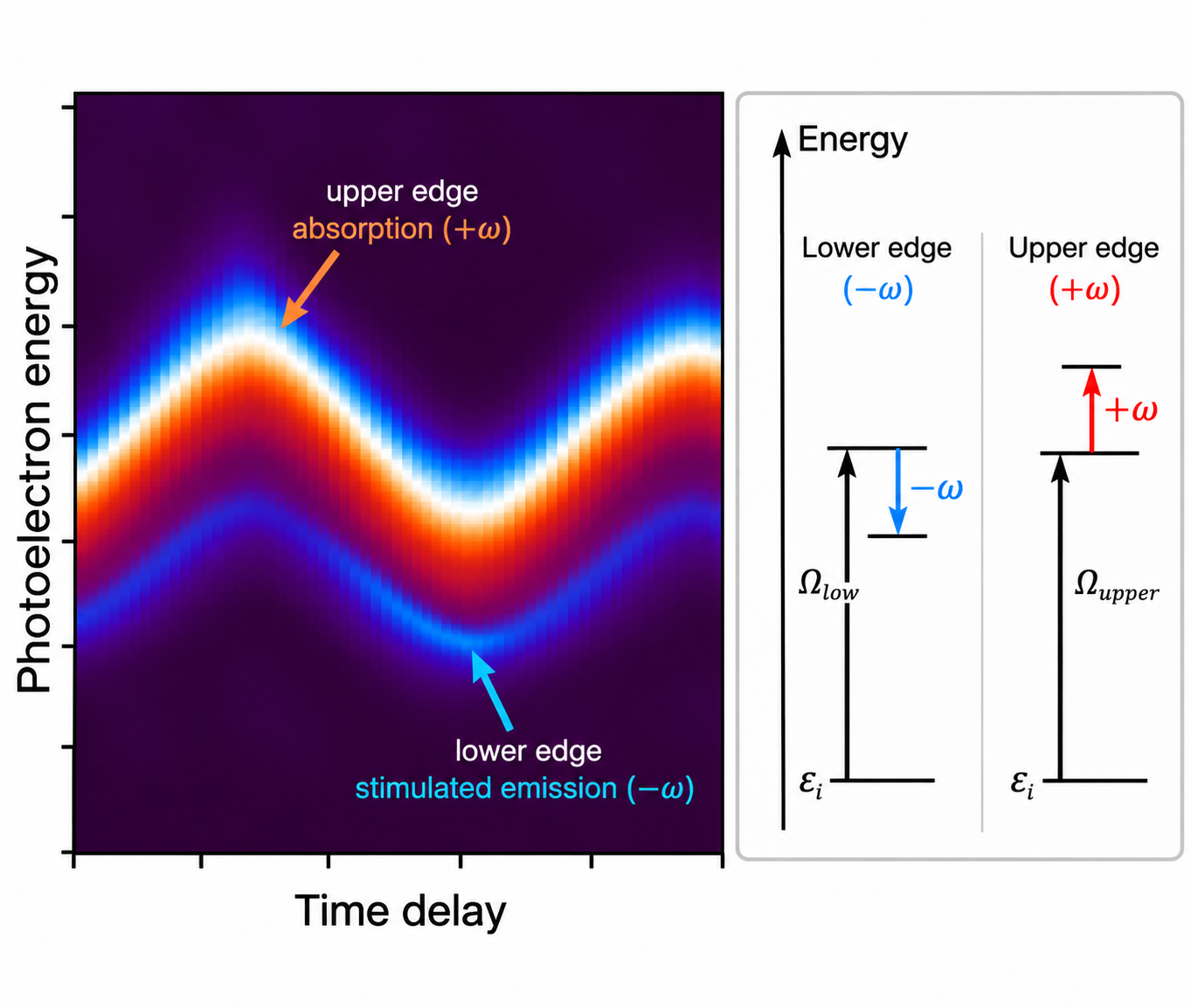}

\caption{\label{fig:scheme} Schematic attosecond streaking spectrogram
  and the corresponding continuum--continuum transition pathways. The
  lower and upper spectral edges are populated by distinct regions of
  the XUV spectrum, characterized by central photon energies
  $\Omega_{\mathrm{low}}$ and $\Omega_{\mathrm{upper}}$,
  respectively. Subsequent stimulated IR emission ($-\omega$)
  down-shifts the photoelectron toward the lower spectral edge,
  whereas IR absorption ($+\omega$) drives it toward the upper
  spectral edge. Adapted from Fig.~4 of ~\cite{Dahlstrom2013}.}
\vs{-5mm}
\end{figure}

At a fixed final photoelectron energy $E$, the fundamental IR
modulation component contains a heterodyne signal born from the
interference between the direct XUV ionization amplitude $D$ and the
two-photon XUV$\pm$IR dressing amplitudes $M$~\cite{Dahlstrom2013}. For
parallel linear laser polarizations and a He $1s$ ground state, these
channels map onto the explicit angular distributions
$D(\theta)=D_1P_1(\cos\theta)$ and
$M^\pm(\theta)=T_0^\pm+2P_2(\cos\theta)T_2^\pm$, structuring the total
fundamental-frequency modulation via
\begin{equation}
 Q_\omega(E,\theta)=D(E,\theta)M^{+*}(E,\theta) + D^*(E,\theta)M^-(E,\theta).
 \label{eq:heterodyne}
\end{equation}
For a broadband isolated attosecond pulse, these pathways are weighted
asymmetrically across the spectrum. As illustrated in
Fig.~\ref{fig:scheme}, XUV photoabsorption from the initial state
$\epsilon_i$ launches intermediate wave packets at kinetic energies
$\epsilon_i+\Omega_{\mathrm{low}}$ and
$\epsilon_i+\Omega_{\mathrm{upper}}$. Subsequent stimulated IR
emission ($-\omega$) down-shifts photoelectrons toward the lower
spectral edge, whereas single-photon absorption ($+\omega$) drives
them toward the upper edge. Near the spectral center, both channels
interfere coherently and are inseparable, 
whereas the spectral edges isolate a clean structure dominated by a
single XUV--IR pathway.

Because the direct XUV and one-IR-photon-assisted amplitudes differ by
a single electric-dipole interaction, they possess opposite parity,
making their fundamental-frequency heterodyne interference strictly
inversion-odd.  Rather than decomposing the raw, instantaneous
photoelectron momentum distribution, we construct the forward-backward
asymmetric component
\begin{equation}
 I_{\rm odd}(E,\theta,\tau)=
 \left[I(E,\theta,\tau)-I(E,\pi-\theta,\tau)\right]/2 \ ,
 \label{eq:fold}
\end{equation}
where $\theta$ denotes the polar angle. We subsequently isolate the
complex Fourier coefficient at the IR fundamental frequency. This
specific sequence of operations eliminates the dominant inversion-even
background before any angular decomposition is performed. Empirically,
we find that the surviving inversion-odd signal is overwhelmingly
dominated by the $P_1+P_3$ Legendre sector across the entire energy
and delay range under investigation.

Expanding the isolated inversion-odd modulation component in a
Legendre series,
\begin{equation}
 C_\omega^{\rm odd}(E,\theta) =A_1(E)P_1(\cos\theta)+A_3(E)P_3(\cos\theta)+\cdots,
 \label{eq:odd}
\end{equation}
and exploiting the recurrence relation $P_1P_2={2/5}P_1+{3/5}P_3$, we
find that a single isolated pathway yields the coefficients
\begin{equation}
 A_1\propto T_0+{4/5}T_2,\qquad A_3\propto {6/5}T_2.
 \label{eq:A13}
\end{equation}
Crucially, this $P_3$ term represents a rank-3 Legendre projection of
lowest-order single-IR-photon $p$--$d$ coherence rather than a
physical $l=3$ partial wave, distinguishing it from the higher-order
multiphoton $f$-wave populations documented in Ref.~\cite{Gao2026NP}.

To quantify the Fano propensity, we define the microscopic
single-IR-photon $s/d$ amplitude ratio as
\begin{equation}
 r^\pm\equiv {T_0^\pm/ T_2^\pm}=R^\pm e^{i\Delta\phi^\pm},
 \label{eq:rdef}
\end{equation}
where $T_0^\pm$ and $T_2^\pm$ denote the $s$- and $d$-wave two-photon
transition amplitudes for IR absorption $(+)$ and emission $(-)$,
respectively. Under this notation, the continuum-continuum Fano
propensity rule dictates the inequality $R^+ < 1 < R^-$. By inverting
the system in Eq.~\eqref{eq:A13}, we map the effective complex $s/d$
ratio directly from the measured Legendre coefficients via
\begin{equation}
\reff(E)=\left[6{A_1(E)/ A_3(E)}-4\right]/5 .
 \label{eq:reff}
\end{equation}
When both the absorption and emission pathways contribute
simultaneously to the signal, the identical algebraic framework yields
the interferometric relation
\begin{equation}
 \reff=[a\,r^{+*}+b\,r^{-}]/(a+b),
 \label{eq:mixed}
\end{equation}
where the weighting coefficients are given by $a=D_1T_2^{+*}$ and
$b=D_1^*T_2^-$. Consequently, the extracted observable converges to
the pure limits $\reff \to r^{+*}$ on an
absorption-dominated spectral edge, and $\reff \to r^{-}$ on a
stimulated-emission-dominated edge. In contrast, the central spectral
region remains a complex coherent interferometer rather than a direct
tracker of individual matrix-element ratios.

\begin{figure}[t]
\centering
\includegraphics[width=1.0\columnwidth]
{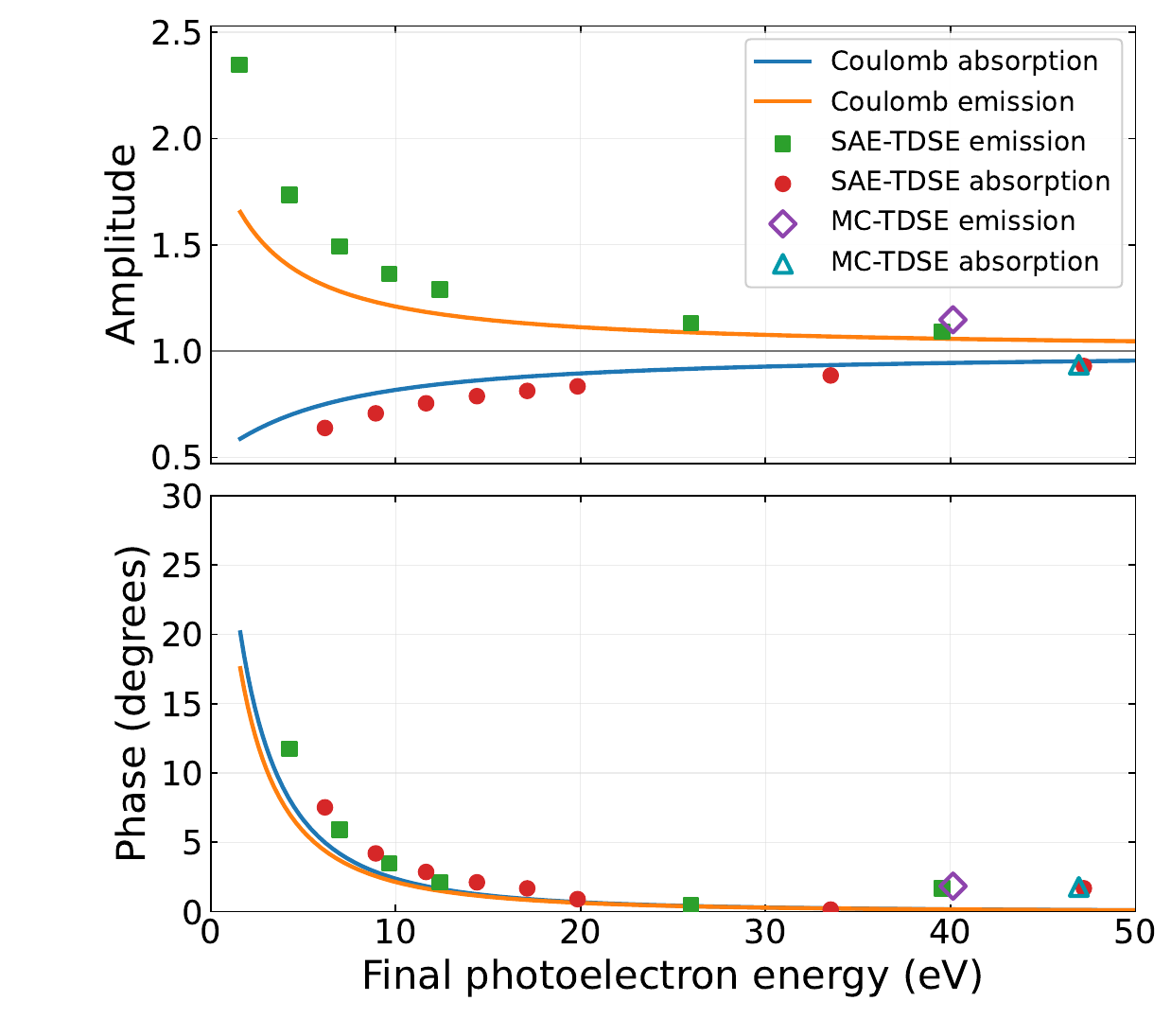}

\vs{-3mm} 
\caption{
\label{fig:continuum} Continuum propensity ratios: amplitude (top) and
phase (bottom). The sign of the emission phase is reversed for compact
presentation.  Analytic Coulomb CC predictions from Ref.~\cite{Ji2024}
compared with TDSE results for 400-as transform-limited XUV
pulses. Both the single active electron (SAE) and multi-channel (MC)
results are shown.  Both TDSE models are described in the End Matter.}
\vs{-5mm}
\end{figure}

We first examine the continuum propensity within a series of benchmark
TDSE simulations driven by a transform-limited 400-as XUV pulse and a
co-propagating IR field with an intensity of $I_{\rm
  IR}=10^{11}~\mathrm{W/cm}^2$. By tuning the XUV carrier frequency
from the near-threshold region up to 2.5~a.u., we systematically
extract the response within isolated spectral windows where either
single-photon IR absorption or stimulated emission channels
dominate. The results of these numerical investigations are summarized
in Fig.~\ref{fig:continuum}.  The upper panel provides a
direct comparison between the effective amplitude ratios extracted
from our TDSE simulations and the analytical,
microscopic Coulomb continuum-continuum transition models formulated
in Ref.~\cite{Ji2024}.

Both the absorption $(+)$ and emission $(-)$ branches obey the Fano
propensity rules, converging smoothly toward unity as the
photoelectron kinetic energy increases. 
The extracted TDSE ratios follow the overall energy dependence of the
Coulomb predictions, with deviations decreasing toward higher
photoelectron energies. The more pronounced differences at low
energies likely reflect short-range non-Coulombic corrections to the
realistic helium ionic potential.

These theoretical predictions are next verified against experimental
helium streaking measurements conducted on the KSU in-line
angle-resolved attosecond beamline~\cite{Gao2026NatCommun}, which is
driven by a carrier-envelope phase-stabilized, few-cycle Yb laser
system. An isolated XUV supercontinuum spanning 60 to 80~eV is
generated via high-harmonic generation in an argon target and
subsequently filtered using thin Al and Zr foils. The XUV and dressing
IR fields are linearly polarized along a common axis, overlapped
collinearly, and focused onto a helium gas target inside a
high-resolution velocity-map imaging (VMI)
spectrometer~\cite{Kling2014VMI}. The VMI spectrometer records the
two-dimensional photoelectron momentum distribution as a function of
the relative XUV-IR time delay. This integrated acquisition preserves
the simultaneously coupled energy, emission-angle, and delay
dependence required for our tomographic analysis. The isolated
attosecond XUV pulse possesses a duration of approximately 150~as with
a positive intrinsic chirp~\cite{Gao2026NP}. Crucially, while the
original analysis in Ref.~\cite{Gao2026NP} relied on direct
projections of the momentum data onto the laser polarization plane,
the present study implements full Abel inversions using both the
BASEX~\cite{Dribinski2002} and rBASEX~\cite{Garcia2004,Ryazanov2012}
algorithms to cleanly reconstruct the true central momentum slice in
that plane.

\begin{figure}[h]
\centering
\includegraphics[width=1.0\columnwidth]
{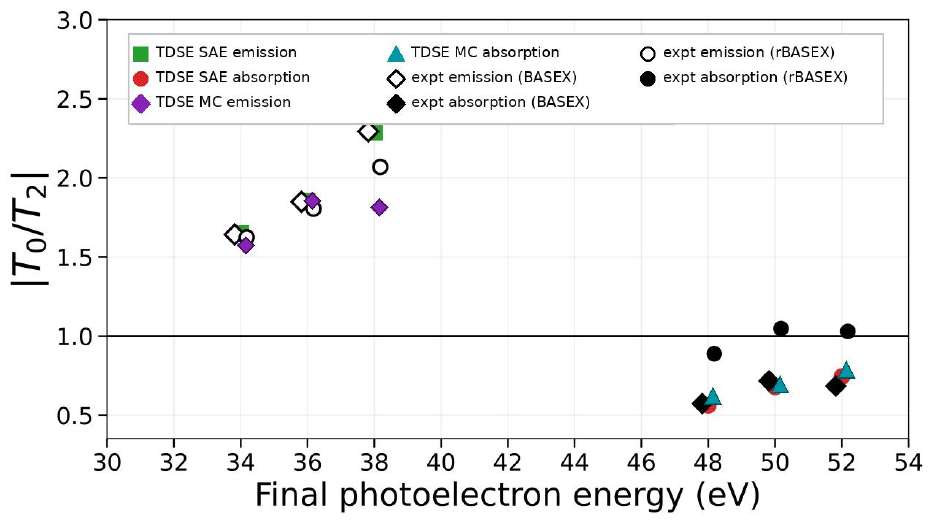}

\vs{-3mm} 
\caption{
\label{fig:chirped} Continuum propensity ratios. 
  Chirped SAE and MC TDSE ratios are compared with experimental BASEX
  and rBASEX reconstructions in 2-eV bins labeled by their lower
  edges.}  \vs{-5mm}
\end{figure}

For a direct comparison with the experiment, we repeat the simulation
at $\Omega=2.56$~a.u. using a 158-as XUV pulse and a laser intensity
of $I_{\rm IR}=4\times10^{11}~\mathrm{W/cm}^2$, incorporating the
experimental XUV attochirp. This quadratic phase shift alters the
relative complex weights of the competing pathways, driving the
extracted ratio farther from unity on the emission side while keeping
the absorption side below unity. These chirped values represent the
effective interferometric ratio $\reff$ from Eq.~\eqref{eq:mixed},
rather than a path-pure microscopic parameter $r^\pm$. To match the
experimental data reduction, we group the TDSE results into identical
2-eV bins and focus exclusively on the extreme lower (34--38~eV) and
upper (48--52~eV) spectral edges where single-pathway selection is
maximized. The central regions (40--46~eV) are omitted because the
strong coherent mixing of absorption and emission amplitudes prevents
independent pathway extraction via Eq.~\eqref{eq:mixed}.

The horizontal separation between the emission and absorption data
points in Figs.~\ref{fig:continuum} and \ref{fig:chirped} reflects their
extraction from distinct, path-selective spectral windows. Their
relative offset is therefore dictated by the finite XUV pulse
bandwidth and the specific selection of the corresponding spectral
edges. This separation should not be confused with the
fixed $2\omega$ spacing between the nominal XUV photon energies of the
absorption and emission pathways that would otherwise lead to an
identical final photoelectron kinetic energy.

The agreement between theory and experiment is particularly robust on
the emission edge, where both BASEX and rBASEX reconstructions closely
follow the rising trend of the chirped SAE-TDSE results. The
five-channel calculation preserves the emission-side Fano propensity,
although its highest-energy ratio is lower than the experimental values. In contrast, the weaker
absorption edge exhibits a higher sensitivity to the choice of
reconstruction method; the chirped TDSE and BASEX results remain below
unity, whereas the rBASEX values hover slightly above unity. Because the
absorption-side Fano inequality cannot be established completely
independent of the inversion technique, we treat the emission branch
as our primary benchmark and retain both inversions in
Fig.~\ref{fig:chirped} to explicitly display this systematic
uncertainty. Ultimately, the robust lower-edge agreement confirms that
the experimental $\reff$ closely follows the predicted chirped TDSE
behavior under realistic broadband-pulse dressing conditions.

Upon approaching the ionization threshold, the familiar semiclassical
picture of attosecond streaking as a mere velocity modulation
governed by the IR vector potential breaks down. For example, previous
work by Xu et al.~\cite{Xu2011} demonstrated that low-energy
photoelectrons subjected to intense dressing fields can undergo
trajectories where the laser field reverses the electronic motion,
driving electron-ion rescattering dynamics that depart qualitatively
from conventional streaking. The regime investigated in our study is
physically distinct; our relatively weak IR field lies orders of
magnitude below that strong-field ionization threshold, ensuring that
such rescattering mechanisms do not dominate. Consequently, the
delay-dependent XUV-IR streaking observable and its fundamental
Fourier modulation component remain strictly well defined. Notably,
as the intermediate photoelectron energy drops below the ionization
threshold, this same frequency-resolved observable transitions to a
regime governed by resonant pathways through the bound discrete He
$np$ Rydberg manifold.

To probe discrete intermediate states, we perform a separate
SAE-TDSE calculation with a transform-limited 400-as XUV pulse
centered at $\Omega=0.90$~a.u. and a weak 204-nm dressing field
($\omega=0.22381$~a.u., $I_{\rm IR}=10^{11}$~W/cm$^2$).
Absorption of one dressing photon maps the bound He $np$ states
onto final photoelectron energies $E_{np}+\omega$, including
2.553~eV for $2p$ and 4.544~eV for $3p$ in our SAE potential.
The broadband XUV spectrum preferentially weights the absorption
pathway in this energy range, bringing the effective ratio closer
to its absorption-path limit in Eq.~\eqref{eq:mixed}.

\begin{figure}[t]
\centering
\includegraphics[width=\columnwidth]{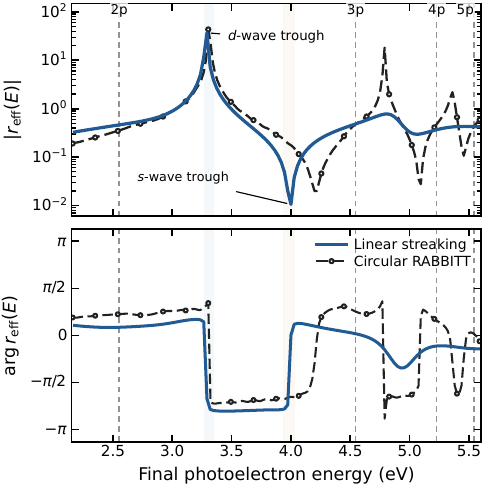}

\caption{\label{fig:resonant_204nm} Effective $s/d$ ratio for helium
  in the resonant UV-dressing regime. The top panel shows the modulus
  $|r_{\rm eff}(E)|$ and the bottom panel the wrapped phase $\arg
  r_{\rm eff}(E)$ extracted from linear streaking for a 400-as XUV
  pulse centered at $\Omega=0.90$ a.u. and a 204-nm dressing
  field. The corresponding circular RABBITT result \cite{Serov2026} is
  overlaid for comparison. 
}
\end{figure}

Figure~\ref{fig:resonant_204nm} shows a pronounced resonant structure
in the effective $s/d$ ratio. In the top panel, $|\reff|$ rises
sharply near 3.29~eV and falls to a deep minimum near 4.00~eV, between
the shifted Rydberg-state energies. These extrema have a simple
partial-wave origin: the absorption-dominated $d$-wave two-photon
amplitude develops a destructive-interference \emph{trough} near
3.29~eV while the $s$-wave amplitude remains finite, producing a large
$s/d$ ratio. Conversely, a displaced trough in the $s$-wave amplitude
near 4.00~eV produces the ratio minimum. Such antiresonances arise
from interference between the $np$-mediated excitation and the
nonresonant continuum, whose relative strengths differ in the $s$ and
$d$ channels~\cite{Serov2026,Drescher2022}.  The rapid phase
variations in the bottom panel accompany these complex-amplitude
cancellations.

For comparison, Fig.~\ref{fig:resonant_204nm} also shows the
counterrotating circular-RABBITT absorption ratio and phase of
Ref.~\cite{Serov2026}. The first prominent ratio maximum closely
coincides with the streaking result, while higher-$np$ structures
differ in detail, reflecting differences in XUV pulse duration. This
correspondence supports a common resonant-antiresonant mechanism
without implying numerical identity; the precise height of the sharp
maximum is sensitive to the near-zero denominator.

To verify the resilience of this isolated structure at higher
intensities, we scale the laser intensity from $I_{\rm IR}=1 \times
10^{11}$ to $8 \times 10^{11}~\mathrm{W/cm}^2$ at
$\Omega=1.5$~a.u. The extracted effective ratios change very little
over this eightfold intensity variation, while the $P_1$ and $P_3$
coefficients track the lowest-order $I^{1/2}$ heterodyne scaling,
whereas the non-linear $P_5$ background grows as $I^{3/2}$.

In summary, we have uncovered a hidden perturbative angular-momentum
structure buried within the intrinsically multiphoton architecture of
attosecond streaking in helium. By pairing forward-backward spatial folding with
Fourier demodulation at the IR fundamental frequency, we isolate a
clean $P_1+P_3$ response whose complex coefficient ratio maps
analytically onto the continuum-continuum Fano propensity. Our
transform-limited TDSE benchmark accurately reproduces the smooth
energy dependence of the Coulomb continuum, while calculations
incorporating the experimental pulse chirp successfully reproduce the
measured trends on the robust emission edge across independent
tomographic inversion methods. Crucially, the filtered response
retains its predominantly lowest-order angular signature even as
higher-order processes heavily populate the full streaking wave
packet. Moving beyond the continuum, our formulation predicts that
sub-threshold excitation uncovers clear amplitude and phase
fingerprints governed by the bound discrete He $np$ Rydberg
series. These findings demonstrate that attosecond streaking encodes
rich quantum information far beyond the classical velocity modulation
of photoelectrons. By isolating the lowest-order XUV--IR interference from the
multiphoton background, this protocol establishes a spectroscopic
pathway to map continuum propensities and discrete-state resonances
in the time domain. 
While demonstrated here explicitly for helium, this symmetry- and
frequency-filtered methodology offers a general framework for
extending the analysis to atomic, molecular, and solid-state systems
where attosecond streaking is an established technique.

\begin{acknowledgments}
The Kansas State University work was supported by the Chemical
Sciences, Geosciences and Biosciences Division, Office of Basic Energy
Sciences, Office of Science, U.S. Department of Energy, under Grant
No.~DE-FG02-86ER13491.  Jingsong~Gao~was supported by the Grant  DE-SC0026268
from the same funding agency. Meng~Han~was supported by the
U.S. National Science Foundation through a Faculty Early Career
Development (CAREER) Award  2543157.

The theoretical work used resources of the
National Computational Infrastructure, Australia.  This work was
supported in part by the Australian Research Council Discovery Project
DP230101253.
\end{acknowledgments}

\clearpage
\section*{End Matter}

\paragraph{TDSE solution and numerical parameters.}

We solve the two-electron TDSE in a five-channel basis using the PAHF
implementation of Ref.~\cite{PhysRevA.84.062701}.
The wave function is represented on a finite-element
discrete-variable radial grid with a spherical-harmonic
expansion. Time propagation employs an implicit
second-order scheme, while exterior complex scaling
suppresses reflections from the radial boundary.
This implementation has also been benchmarked in
earlier attosecond calculations
\cite{Serov2026,Serov2026rainbow}.
For the continuum calculations in Fig.~\ref{fig:continuum},
the IR pulse has a 5.0-fs intensity FWHM and carrier frequency
$\omega=0.04796$~a.u. (950~nm), with 13 delays
$\tau=-120,-100,\ldots,120$~a.u. (0.484-fs spacing). The
resonant SAE calculation in Fig.~\ref{fig:resonant_204nm} instead
uses a 400-as XUV pulse at $\Omega=0.90$~a.u. and a
204-nm dressing field ($\omega=0.22381$~a.u.,
$I_{\rm IR}=10^{11}$~W/cm$^2$), sampled at 31 delays
$\tau=-120,-112,\ldots,120$~a.u. to resolve the shorter UV optical period. The inversion-odd signal
is demodulated at the physical dressing frequency in each case.
Both theoretical delay windows span about $\pm2.90$~fs, shorter
than the experimental $\pm10$-fs window. Closely spaced higher-$n$
features may depend on the finite XUV bandwidth and delay sampling;
convergence of their detailed line shapes is not claimed.

\paragraph{Experimental demodulation and angular analysis.}
The experimental archive contains 15 matrices
$I(E,\alpha,\tau)$ labelled $E=30,32,\ldots,58$~eV.  The BASEX and
rBASEX central-slice reconstructions used in this analysis contain the
corresponding matrices from 30 to 52~eV.  In all cases the label denotes
the lower edge of a hard 2-eV top-hat energy bin.  Each matrix contains
360 detector-angle rows and 121 delays.  The delay interval is
$-10\le\tau\le10$~fs with a 0.167-fs step.  At each detector angle an
independent constant and linear temporal background is removed before
Hann-windowed demodulation,
\begin{equation}
 C_\omega(E,\alpha)=
 {\sum_j w_j[I(E,\alpha,\tau_j)-a-b\tau_j]e^{-i2\pi f_0\tau_j}
 \over\sum_jw_j},
\end{equation}
with $f_0=0.302$ fs$^{-1}$ ($T_0=3.311$ fs).
The experimental IR pulse is broadband (approximately
600--1250~nm), with a nominal central wavelength near
1000~nm. The TDSE calculations use a 950-nm carrier,
while experimental demodulation uses $f_0$, corresponding
to approximately 993~nm.
The inversion components are
\begin{equation}
 C_\omega^{\rm odd/even}(\alpha)=
 {C_\omega(\alpha)\mp C_\omega(\alpha+180^\circ)\over2}.
\end{equation}
The odd component is then folded to the physical polar angle after averaging the two equivalent detector directions related by $p_x\to-p_x$.

The complex Legendre coefficients are defined by
\begin{equation}
 A_L(E)={2L+1\over2}\int_0^\pi
 C_\omega^{\rm odd}(E,\theta)P_L(\cos\theta)\sin\theta\,\dd\theta,
\end{equation}
for odd $L$. The selected $P_1+P_3$ sector dominates the odd-Legendre signal throughout the analyzed range.

\paragraph{Exact mixed-path relation.}
With $D=D_1P_1$ and $M^\pm=T_2^\pm[2P_2+r^\pm]$, the coefficient of the fundamental is
\begin{equation}
 Q=P_1\{a[2P_2+r^{+*}]+b[2P_2+r^{-}]\},
\end{equation}
where $a=D_1T_2^{+*}$ and $b=D_1^*T_2^-$.  Using $2P_1P_2=(4/5)P_1+(6/5)P_3$ gives
\begin{align}
 A_1&=a\left(r^{+*}+{4\over5}\right)
      +b\left(r^{-}+{4\over5}\right),\\
 A_3&={6\over5}(a+b),
\end{align}
and hence Eq.~(\ref{eq:mixed}).  The formula makes both the power and the limitation of the method explicit.  On a path-pure edge it gives a microscopic CC ratio, whereas in a mixed region a small $a+b$ can amplify $A_1/A_3$ without requiring an anomalously large individual matrix-element ratio.

\paragraph{Robustness of the experimental extraction.}
Varying the fundamental frequency over 0.292--0.312~fs$^{-1}$,
replacing full-window Hann demodulation by fixed-frequency harmonic
regression, and varying the angular fitting range preserve the
emission-edge ratios.  The central slice was reconstructed
independently with BASEX and rBASEX.  BASEX has a known numerical
artifact close to $0^\circ$ and $180^\circ$, while rBASEX largely
removes this feature but is less continuous close to $90^\circ$.
Removing narrow angular caps around the affected directions leaves the
34--38~eV emission ratios essentially unchanged.  The two inversions
therefore give a consistent emission branch.  Their difference is
larger on the weak absorption bins labelled 48--52~eV (covering
48--54~eV): BASEX remains below
unity, whereas rBASEX reaches and slightly exceeds unity at 50 and
52~eV.  We therefore do not use the absorption edge alone as an
inversion-independent test of the Fano inequality.  Both data sets are
retained in Fig.~\ref{fig:chirped} to display this reconstruction
systematic explicitly.  The experimental comparison in
Fig.~\ref{fig:chirped} is based on this single published streaking
data set, analyzed with the two independent central-slice inversions.

\paragraph{Doubly excited helium states.}
To assess two-electron effects in the spectral region
where doubly excited helium states are accessible, we
perform five-channel TDSE calculations including the
He$^+ (1s, 2s, 2p)$ ionic states. At lower photon
energies, we employ the single-active-electron (SAE)
approximation. Comparisons of the SAE and multichannel
calculations, for unchirped and chirped parameter sets,
show that the extracted Fano propensity is largely
insensitive to two-electron corrections. 
Table~\ref{tab:sae-mc-ratios} gives the corresponding edge ratios.
The largest multichannel correction, approximately 22\%, occurs
in a chirped emission bin without reversing the Fano propensity.

\begin{table}[t]
\caption{\label{tab:sae-mc-ratios} Single-active-electron (SAE) and
  five-channel (MC) TDSE effective amplitude ratios $|r_{\rm eff}|$
  from the inversion-odd IR fundamental.  For the unchirped
  $\Omega=2.50$~a.u. calculation (400-as XUV, $I_{\rm
    IR}=10^{11}$~W/cm$^2$), entries are signal-weighted averages of
  energy-resolved ratios over identical final-energy windows in both
  models. The MC symbols in Fig.~\ref{fig:continuum}(top) were
  extracted independently from the complex amplitude matrices using
  $|A_3|$-weighted averages and therefore differ slightly from the
  values tabulated here.  The SAE symbols denote the original
  energy-local benchmark sweep.  For the chirped
  $\Omega=2.561$~a.u. calculation (158-as XUV, $\beta=-112$~fs$^{-2}$,
  $I_{\rm IR}=4\times10^{11}$~W/cm$^2$), entries are effective ratios
  obtained from the 2-eV-binned angular signal.  The two parameter
  sets are distinct, rather than a chirp-only comparison.  MC
  residual-ion channel probabilities are summed incoherently.}
\begin{ruledtabular}
\begin{tabular}{lccc}
Path & $E_{\rm kin}$ & SAE & MC \\
\hline
\multicolumn{4}{c}{Unchirped $\Omega=2.50$~a.u.} \\
Emission & $1.40$--$1.55$~a.u. & 1.301 & 1.215 \\
Absorption & $1.65$--$1.80$~a.u. & 0.965 & 0.984 \\
\hline
\multicolumn{4}{c}{Chirped $\Omega=2.561$~a.u.} \\
Emission & $34$--$36$~eV & 1.667 & 1.569 \\
Emission & $36$--$38$~eV & 1.878 & 1.855 \\
Emission & $38$--$40$~eV & 2.318 & 1.812 \\
Absorption & $48$--$50$~eV & 0.585 & 0.620 \\
Absorption & $50$--$52$~eV & 0.701 & 0.694 \\
Absorption & $52$--$54$~eV & 0.746 & 0.786 \\
\end{tabular}
\end{ruledtabular}
\end{table}

\paragraph{Perturbative-to-nonperturbative continuation.}
For the independent $\Omega=1.5$ a.u. TDSE series, the same 13-delay grid spanning approximately two optical cycles is analyzed at $I_{\rm IR}=1$, 4, and $8\times10^{11}$ W/cm$^2$, with all other pulse parameters held fixed across the series.  The quoted ratios are medians over fixed windows $0.37<E<0.47$ a.u. (emission) and $0.68<E<0.78$ a.u. (absorption), giving 1.510, 1.570, and 1.660 and 0.756, 0.767, and 0.772, respectively.  For the intensity scaling, $|A_L|$ is first geometrically averaged over these two edge windows; the resulting amplitudes obey $|A_L|\propto I^{\alpha_L}$ with $\alpha_1=0.474$, $\alpha_3=0.456$, and $\alpha_5=1.365$.  Meanwhile the population carried by high angular momenta in the complete wave packet grows strongly with intensity.  The selected odd fundamental therefore remains close to the perturbative angular structure even when the complete state does not.

\paragraph{Relation to the published angle-resolved streaking benchmark.}
Ref.~\cite{Gao2026NP} used the same experimental acquisition to
characterize multiphoton angular-momentum transfer in the full
streaking wave packet, reconstruct the experimental XUV and IR fields,
and benchmark the measured energy- and angle-resolved traces against
SAE-TDSE and SFA calculations.  The present Letter asks a different
question: whether a symmetry- and frequency-selected component of that
broadband multiphoton signal can be mapped onto the elementary CC
propensity.  We therefore isolate the IR fundamental, separate its
inversion-odd component, project it onto $P_1$ and $P_3$, and use their
complex ratio to construct $r_{\rm eff}$.  The 400-as transform-limited
TDSE sweep then provides a clean continuum benchmark against the
Coulomb CC amplitudes, while the experimental comparison uses a 158-as
pulse with the experimentally determined XUV attochirp,
$\beta=-112$~fs$^{-2}$ in the TDSE field convention, analyzed in the
same 2-eV bins as the BASEX and rBASEX central slices.  In this sense
the experimental comparison tests the survival of a propensity-sensitive
observable under the measured broadband-pulse conditions; it does not
identify the measured $r_{\rm eff}$ numerically with a bare $r^\pm$.
The continuation of this selected observable into the discrete Rydberg
regime is an additional theoretical prediction not considered in
Ref.~\cite{Gao2026NP}.

\paragraph{Data and code availability.}
The experimental data underlying the streaking measurements are publicly
available via Zenodo~\cite{Han2026ZenodoData}. 
The theoretical data and analysis scripts are publicly available on
Zenodo at https://doi.org/10.5281/zenodo.23515125.

\end{document}